\documentclass[prb,showpacs,amsmath,amssymb,twocolumn,floatfix]{revtex4}
\usepackage{amsfonts}
\usepackage{times}
\usepackage{graphicx}
\usepackage{dcolumn}
\usepackage{bm}
\begin{document}
 
\title{A Continuum, ${\mathcal O}(N)$ Monte-Carlo algorithm for charged
particles}
\author{J\"org Rottler}
\email{joerg@turner.pct.espci.fr}
\author{A. C. Maggs}
\email{tony@turner.pct.espci.fr}
\affiliation{Laboratoire de Physico-Chimie Th\'eorique, 
UMR CNRS-ESPCI 7083, 10 rue Vauquelin, F-75231 Paris Cedex 05, France}
\date{\today}
\begin{abstract} 
We introduce a Monte-Carlo algorithm for the simulation of charged
particles moving in the continuum. Electrostatic interactions are not
instantaneous as in conventional approaches, but are mediated by a
constrained, diffusing electric field on an interpolating lattice. We
discuss the theoretical justifications of the algorithm and show that
it efficiently equilibrates model polyelectrolytes and polar fluids.
In order to reduce lattice artifacts that arise from the interpolation
of charges to the grid we implement a local, dynamic subtraction
algorithm. This dynamic scheme is completely general and can also be
used with other Coulomb codes, such as multigrid based methods.
\end{abstract}
\maketitle

\section{Introduction}
Simulating charged condensed matter systems is demanding due to the
long range nature of the Coulomb interaction \cite{schlick,sagui}. The
direct evaluation of the Coulomb sum for $N$ particles,
$U_c=\sum_{i<j}e_ie_j/4\pi\epsilon_0 r_{ij}$, requires computation of
the separations $r_{ij}$ between all pairs of particles, which implies
${\mathcal O}(N^2)$ operations are needed per sweep or time step.
Conventional fast algorithms including the classic Ewald sum
\cite{ewald32}, particle mesh Ewald approaches \cite{ewald,ppm}, and
fast multipole algorithms \cite{multipole} suffer from either poor
scaling with system size, high coding complexity or inefficiency in
multiprocessor environments.  Although multigrid methods \cite{darden}
appear to be a promising candidate for fast parallelized molecular
dynamics simulations, these techniques are less efficient for
Monte-Carlo simulations, where the total energy of the system must be
recalculated after {\em every} particle move.

It is surprising that the Coulomb interaction poses such tremendous
difficulty; after all the underlying Maxwell equations are {\em
local}. The above methods, however, do not solve Maxwell's equations,
but rather search for the electrostatic potential $\phi_p({\bf r})$,
from which the electric field $\bf{E(r)}$ is deduced.  We shall show
in this article that this full relaxation of the electric field to the
solution of the Poisson equation is not necessary in the context of
statistical mechanics, where one is interested in configurational
averages. This allows us to construct a nonstandard approach that is
{\em local} and substantially more efficient than methods involving
potential theory.

Recently, a lattice Monte-Carlo algorithm was introduced in which the
time needed for a single sweep scales as only ${\mathcal O}(N)$.  This
algorithm employs the electric field, ${\bf E}$, rather than the
scalar potential as the true degree of freedom \cite{acm,acm2}.  For a
charge density $\rho({\bf r})=\sum_ie_i\delta({\bf r}-{\bf r}_i)$,
Gauss' law, $\epsilon_0{\rm div}\,{\bf E}({\bf r})-\rho({\bf r})=0$,
is imposed as an exact dynamical constraint on the field
configurations.  The algorithm correctly reproduced the static
properties of a charged lattice gas and equilibrated the system with a
${\mathcal O}(1)$ overhead when compared to the time needed to
simulate a neutral system \cite{acm3}. This high efficiency results
from purely local updates of particles and fields using the standard
Metropolis Monte-Carlo rule; there is no global calculation of the
potential.

In the present article, we generalize this algorithm to off-lattice
Monte-Carlo, so that the particles move in the continuum. Such a
generalization is necessary for any realistic modeling of charged
systems, such as polyelectrolytes, biopolymers and polar solvents.  We
begin with a complete description of the algorithm in Section
\ref{alg-sec} and continue with tests of the static and dynamic
behavior in Section \ref{res-sec}. In particular, we demonstrate
numerically that the slowest relaxation time in the charged system is
dominated by diffusion of the density degrees of freedom, in a manner
which is very similar to a neutral system with short range
interactions.  The following Section \ref{yuk-sec} discusses the
elimination of lattice artifacts that arise from the interpolation of
charges to a grid via dynamic subtraction.  Although demonstrated in
the context of our auxiliary field Monte-Carlo algorithm, this dynamic
subtraction can also be used with more conventional Coulomb algorithms
such as multigrid.

\section{Coulomb interactions from auxiliary fields}
\label{alg-sec}
\subsection{Constrained energy functional}
The starting point for our derivation of a local Coulomb algorithm is
a formulation of electrostatics in terms of a constrained energy
functional based on a vector field $\bf{E(r)}$ \cite{schwinger},
\begin{equation}
{\mathcal F}[{\bf E}]=\int \left[\frac{\epsilon_0{\bf E}({\bf
r})^2}{2}-\phi({\bf r})(\epsilon_0{\rm div\,}{\bf E}({\bf r})-\rho({\bf
r}))\right]d^3r.
\label{func-eq}
\end{equation}
In this functional Gauss' law is imposed with the Lagrange multiplier
$\phi$ and constrains the longitudinal component of the electric
field; $\phi({\bf r})$ can be identified with the scalar potential
$\phi_p$.  Minimization of the functional with respect to ${\bf E}$
leads immediately to the familiar equations of electrostatics,
\begin{eqnarray} 
\label{field-eq}
 {\bf E} = -  \nabla \phi_p\\
\nabla^2 \phi_p = -\rho/\epsilon_0.
\label{poiss-eq}
\end{eqnarray}

All conventional Coulomb algorithms are essentially concerned with
finding solutions to Eqs.~(\ref{field-eq},\ref{poiss-eq}).  After
inserting Eqs.~(\ref{field-eq},\ref{poiss-eq}) into Eq.~(\ref{func-eq}),
we obtain the electrostatic energy $U_c=\frac{\epsilon_0}{2}\int
(\nabla\phi_p)^2 d^3r$ as a minimum of the constrained
functional. Note that a scalar functional based only on the potential
$\phi$ cannot be used, it leads to the wrong sign in the Coulomb
interaction (see Ref.~\onlinecite{acm3} for a discussion of this
point).

The crucial point of our algorithm is now not to minimize the
constrained energy functional Eq.~(\ref{func-eq}), but rather to
maintain the electric field at finite temperature. For $T>0$, we have
\begin{equation}
{\bf E} = -  \nabla \phi_p+{\bf E}_{tr},
\end{equation}
where ${\bf E}_{tr}=\rm {curl}\, {\bf Q}$ is an arbitrary rotational
vector field. The total field ${\bf E}$ still satisfies Gauss' law,
but the energy $U$ is no longer equal to $U_c$, but rather
$U=\frac{\epsilon_0}{2} \int \left \{(\nabla\phi_p)^2 +{\bf E}_{tr}^2
\right \}\; d^3r$. We now show that an integration over these
auxiliary transverse (rotational) degrees of freedom of the electric
field nevertheless lead to electrostatic interactions.

To this end, let us consider the partial partition function for
charged particles and a fluctuating electric field that is constrained
by Gauss' law,
\begin{equation}
{\cal Z}(\{{\bf r}_i\}) = \int {\cal D}{\bf E}\,\prod_{\bf r}\delta({\rm
div} {\bf E} -\rho(\{{\bf r}_i\})/\epsilon_0) e^{-\int d^3r'\, \epsilon_0
{\bf E}^2/2k_BT}.
\end{equation}
Note that the integration is only performed over field configurations
and not particle positions. We evaluate the integral by changing
variables, ${\bf E}_{tr}={\bf E}+\nabla \phi_p$,
\begin{eqnarray}
{\cal Z}(\{{\bf r}_i\}) & = & \int {\cal D}{\bf
 E}_{tr}\prod_r\delta({\rm div\,} {\bf E}_{tr})\, e^{-\int d^3r'\,
 \epsilon_0 ({\bf E}_{tr}-\nabla \phi_p)^2/2k_BT}\nonumber\\ &= &
 e^{-\int d^3r' \epsilon_0 (\nabla \phi_p)^2/2k_BT} \nonumber\\
 &\times& \int {\cal D}{\bf E}_{tr}\prod_{\bf r}\delta({\rm div\,}
 {\bf E}_{tr})\,e^{-\int d^3r'\, \epsilon_0 {\bf
 E}_{tr}^2/2k_BT}\nonumber\\ &=& {\cal Z}_{Coulomb}(\{{\bf
 r}_i\})\times {\rm const.}
\label{part-eq}
\end{eqnarray}
The cross term in the energy, $\int {\bf E}_{tr}\nabla \phi_p d^3r$,
vanishes for periodic or constant potential boundary conditions, as
can be seen by integration by parts. Eq.~(\ref{part-eq}) shows that an
integration over the transverse modes merely multiplies the coulombic
partition function ${\cal Z}_{Coulomb}$ by a constant. Since this
multiplication does not change the relative statistical weight of
configurations, it can be ignored.

This result allows us to maintain the total electric field at finite
temperature rather than quenching it to zero as when solving Poisson's
equation. As long as the constraint of Coulomb's law is strictly
maintained, an integration over all possible transverse field
components will generate an effective Coulomb interaction between the
particles. As illustrated below, this integration involves only local
operations and requires no global optimization. Although this may
sound unusual, it is actually a very natural way of calculating the
interactions: we shall see below that the resulting dynamics has a
local conservation law exactly as occurs in Maxwell's equations.

\subsection{Algorithm}
Our Monte-Carlo algorithm evaluates the partition function for a
Coulomb system using the Metropolis criterion together with the energy
functional Eq.~(\ref{func-eq}). The algorithm consists of two
Monte-Carlo moves that
\begin{itemize}
\item integrate over the particle positions by displacing the charged
particles, but maintain a field configuration that {\em always}
satisfies Gauss' law;
\item integrate over the transverse components of the electric field
${\bf E}_{tr}$ to evaluate the configurational integral
Eq.~(\ref{part-eq}).

\end{itemize}
We shall now describe the implementation of the algorithm for the
off-lattice case. A similar description of the algorithm for lattice
simulations has already been given in Ref.~\onlinecite{acm3}. The reader who
is only interested in the results of the algorithm may jump directly
to Section \ref{res-sec}.

\subsubsection{Discretization}
Particles with charge $e_i=\pm 1$ move continuously in a cubic,
periodic simulation cell of volume $L^3$. The electric field ${\bf E}$
is discretized onto a cubic grid of mesh size $a$, and the Cartesian
components of ${\bf E}$ are associated with the $3L^3/a^3$ links of
the lattice. A natural unit of temperature or energy is $T^\star\equiv
e^2/4\pi\epsilon_0 k_B a$. In these units, the Bjerrum length is
$l_b=a T^\star/T$. A typical value in water at room temperature is
$l_b\sim 7\AA$ for monovalent ions.
Charges are interpolated onto the $L^3/a^3$ nodes of the
lattice. Various choices are possible for the interpolation scheme: we
desire a reasonably smooth interpolation and choose a 3rd order scheme
using B-splines \cite{darden,darden2}. In each Cartesian direction,
the charge is distributed onto the 3 mesh points closest to the
particle. Explicitly, the one-dimensional weights are
\begin{eqnarray}
W_-([x])&=&([x]-0.5)^2/2\nonumber\\
W_0([x])&=&0.75-[x]^2\nonumber\\
\label{chargeint-eq}
W_+([x])&=&([x]+0.5)^2/2,
\end{eqnarray}
where $[x]$ denotes the distance between the particle and the nearest
lattice point in units of the lattice spacing. The three-dimensional
weights are constructed by multiplying the three one-dimensional
weights. Our method is not tied to a particular assignment scheme;
higher or lower order schemes can be employed as long as they conserve
charge exactly.

\subsubsection{Particle motion and Hamiltonian paths}
The simulation begins with a field configuration that satisfies the
discretized, integral version of Gauss' law,
\begin{equation}
a^2\sum_j E_{i,j}=e_i/\epsilon_0
\label{intgauss-eq}
\end{equation} 
In this expression $E_{i,j}=-E_{j,i}$ denotes a field component
leaving site $i$ towards site $j$, and the sum is performed over all
links connecting to site $i$; it corresponds to the total flux leaving
site $i$. The success of the algorithm depends on maintaining the
constraint at all times; the field configuration is updated each time
a particle moves.  For a lattice gas, such an update is easy to
perform \cite{acm}: when a particle moves from site $i$ to site $j$,
one changes the field on the traversed link to $E_{i,j}\rightarrow
E_{i,j}\mp e/(\epsilon_0a^2)$ (the sign depends on the direction in
which the particle moves). The resulting field configuration once
again satisfies Eq.~(\ref{intgauss-eq}). The situation is more
complicated in the off-lattice case; a single particle move changes
the charges on $3^3=27$ nodes, more if the particle changes cells.
\begin{figure}
\includegraphics[width=7cm]{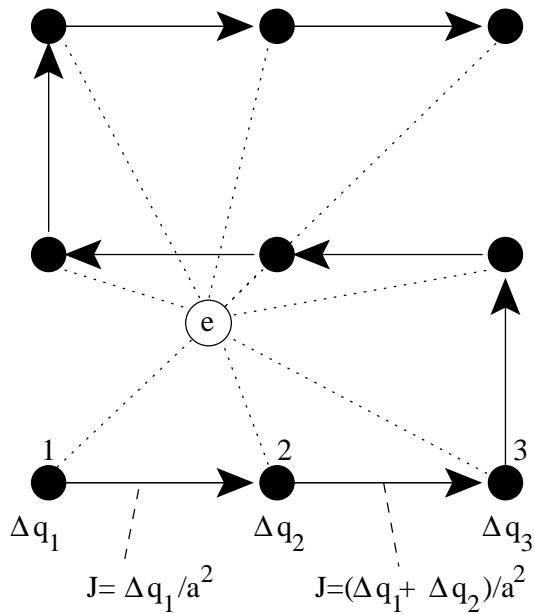}
\caption{\label{path-fig}Illustration of charge interpolation and the
current update using a Hamiltonian path in two dimensions. A charge
(open circle) interpolates onto the $3^2=9$ lattice sites (solid). The
partial charges on these sites change by an amount $\Delta q_i$ when
the particle moves. The current $J$ flows along the Hamiltonian path
and terminates at the last site, since $\sum_i\Delta q_i=0$.}
\end{figure}
One can choose among many update rules to realign the field
configuration at all modified nodes (the system of equations is in
general under-specified). This task can be accomplished with minimal
effort using a path that traverses the modified region in such a way
that it visits each site once. An example of such a path, which is
often called a ``Hamiltonian path'', is shown for two dimensions in
Fig.~\ref{path-fig}; the generalization to three dimensions is
obvious.  Only links along the path need be updated.

The field $E_{i,j}$ on a link that is part of the path is updated
according to $E_{i,j}\rightarrow E_{i,j}\pm \sum_s^\prime \Delta
q_s/(\epsilon_0a^2)$, where $\Delta q_s$ denotes the change of the
fractional charge on a node due to the particle motion. The primed
summation extends over all sites that were previously visited by the
path to reach site $i$, and the sign depends on the direction of the
path. This update is efficiently performed during a single ``walk''
along the path. In a continuum description, the current that flows due
to the motion of the particles produces a change in the electric
field,
\begin{equation}
\label{current-eq}
\frac{\partial {\bf E}}{\partial t}=-{\bf J}/\epsilon_0.
\end{equation}
Note that though the path changes direction several times during the
update, the resulting currents are essentially parallel.  We always
construct the Hamiltonian path in a manner which is symmetric between
the new and old interpolations sites. Because of this our procedure is
time reversible and the algorithm respects detailed balance.  To
simplify the implementation of the algorithm we always displace
particles parallel to the links of the mesh with a step size uniformly
distributed between $0$ and $a$. 

\begin{figure}
\includegraphics[width=8cm]{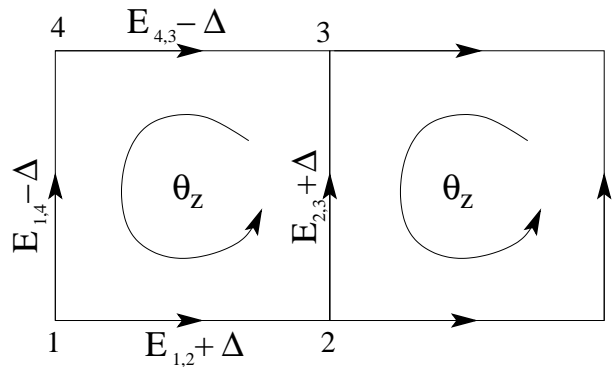}
\caption{\label{plaq-fig}Integration over the transverse degrees of
freedom. The four link fields forming a plaquette in the $xy$-plane
are modified by $\pm \Delta$ in such a way that the total field
changes by a pure rotation $({\rm curl}\,{\bf E})_z=4\Delta$. $\Delta$
is chosen from a uniform distribution. We associate an angular
variable $\Theta_z$ with the plaquette so that $\delta{\bf E}={\rm
curl}\,\delta{\bf \Theta}$.}
\end{figure}

\subsubsection{Integration over the transverse field}
Field updates as described in the previous paragraph change not only
the longitudinal component of the electric field, but simultaneously
affect the transverse field. The particle motion thus already performs
a partial integration over the transverse modes. A complete evaluation
of the configurational integral Eq.~(\ref{part-eq}) often requires a
second type of Monte-Carlo move that changes just the transverse
degrees of freedom of ${\bf E}$ without moving a particle nor
violating the constraint Eq.~(\ref{intgauss-eq}). This is done by
grouping the links that form a faces of the grid into $3L^3$
plaquettes. We randomly choose a plaquette and increment the field on
two of the links by an increment $\Delta$ while decreasing it on the
other two; we only modify the total field configuration by a pure
rotation (see Fig.~\ref{plaq-fig}). We describe the resulting dynamics
by associating an angular variable $\theta_k$ with a plaquette in the
$\{i,j\}$-plane.  As one can see by inspection of Fig.~\ref{plaq-fig},
changes in the electric field are related to the curl of the changes
of the plaquette variable $\theta_k$. Combining this with
Eq.~(\ref{current-eq}), the evolution of the electric field obeys the
equation
\begin{equation}
\label{fulledot-eq}
\frac{\partial {\bf E}}{\partial t}=-{\bf J}/\epsilon_0+{\rm
curl}\,\frac{\partial{\vec \theta}}{\partial t}.
\end{equation}

Changes in ${\bf E}$ also couple back to the circulatory degrees of
freedom, one can show \cite{acm2} that ${\rm curl}\, {\bf E}$ acts
like a torque on ${\vec \theta}$, i.e.
\begin{equation}
\label{thetadot-eq}
\xi\frac{\partial {\vec \theta}}{\partial t}=-\epsilon_0\,{\rm curl}\,
{\bf E}+\vec \eta,
\end{equation}
where we have introduced a relaxation time scale $\xi$ which is
determined by the Monte-Carlo moves; $\vec \eta$ is a Brownian noise.

The dynamics of Eqs.~(\ref{fulledot-eq},\ref{thetadot-eq}) is closely
related to the Maxwell equations \cite{schwinger}. After identifying
$\partial {\vec \theta}/\partial t$ with the magnetic field ${\bf
B}/\mu_0$, we recover Ampere's law in Eq.~(\ref{fulledot-eq}). If we
perform the same replacement in Eq.~(\ref{thetadot-eq}), we recognize
its closeness to another Maxwell equation, Faraday's law. However, the
usual $\partial{\bf B}/\partial t$ has been replaced by just ${\bf
B}$; a time derivative has been lost.  By combining the two equations
one finds \cite{acm2} that the electric field obeys a diffusion
equation
\begin{equation}
\label{diff-eq}
\frac{\partial {\bf E}}{\partial t}=(\epsilon_0\nabla^2{\bf
E}-\nabla\rho)/\xi-{\bf J}/\epsilon_0 +{\rm curl}\,\vec \eta.
\end{equation}
Since in Monte-Carlo we are interested in distributions rather than
dynamics we are free to modify the underlying dynamics. Gauss' law is
imposed as an initial condition and locally conserved both in
Maxwell's equations and in our algorithm, as can be seen by taking the
divergence of Eq.~(\ref{diff-eq}). The conservation of Gauss' law in
our algorithm and in Maxwell's equation is
the origin of Coulomb's law.

\subsection{Boundary Conditions}
Traditional treatments of the Coulomb interaction in periodic systems
lead to a residual dependence on the dielectric constant $\epsilon'$
of a surrounding medium \cite{leeuw1,leeuw2}.  For a standard Ewald
summation, the short-range expansion of the pair potential between two
point charges of opposite sign separated by $r$ is \cite{fraser}
\begin{equation}
V(r)={-e^2 \over 4\pi\epsilon_0 r}-{e^2r^2\over 3\epsilon_0L^3}
\left( {\epsilon'-1\over 2\epsilon'+1}\right)
+{\cal O}(r^4)
\end{equation}
The present algorithm imposes a different choice that we denote
``Maxwell'' boundary conditions, since they are the same as those
generated by the full Maxwell equations.  We show in the appendix that
\begin{equation}
V(r)=
\begin{array}{c}
{-e^2 \over 4\pi\epsilon_0 r}+{e^2r^2\over 3\epsilon_0L^3}
+{\cal O}(r^4)\quad {\rm for } \quad l_b\gg L\\
{-e^2 \over 4\pi\epsilon_0 r}-{e^2r^2\over 6\epsilon_0L^3}
+{\cal O}(r^4)\quad {\rm for }\quad l_b\ll L\label{ew-eq}
\end{array}
.
\end{equation}
There is again a quadratic correction to the bare $1/r$ interaction,
which, surprisingly, depends on the temperature. A crossover occurs
when the Bjerrum length $l_b$ reaches the system size $L$.  It is
possible to remove the temperature dependence of the potential by
introducing a third type of Monte-Carlo move that couples to the mean
or ${\bf q}=0$ component of the electric field $\bar{\bf E}={1\over
L^3}\int d^3r\, {\bf E({\bf r})}$.  The potential then coincides with
the conventional Ewald sum with ``tinfoil'' $(\epsilon'=\infty)$
boundary conditions.

\section{Results}
\label{res-sec}

\subsection{Static quantities}
\begin{figure}
\includegraphics[width=8cm]{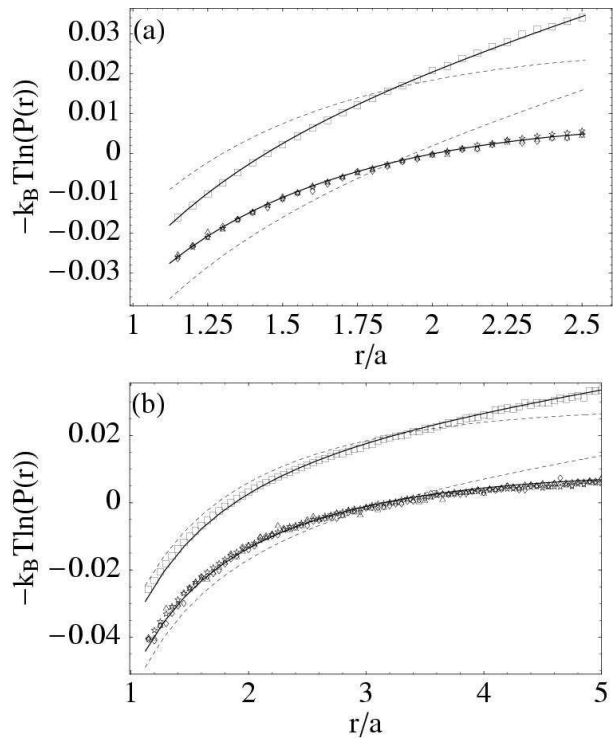}
\caption{\label{pot-fig}Pair potentials obtained from the distribution
$P(r)$ at distance $r$ of a pair of charges for two system sizes (a)
$L=5\,a$ and (b) $L=10\,a$ and two different temperatures
$T_l=0.065T^\star$ ($\square,\star$) and $T_h=1T^\star$
($\triangle,\lozenge$).  Solid lines Eq.~(\ref{ew-eq}), $V(r)={\rm
const}- 1/4\pi r- b(T)r^2$, where the limiting values for the
quadratic correction are $b(T_l)=-1/3\epsilon_0L^3$ and
$b(T_h)=1/6\epsilon_0L^3$ for Maxwell boundary conditions
$(\square,\triangle)$ (dashed lines show the curves in the opposite
limit to demonstrate sensitivity). Addition of a third move leads to
tinfoil boundary conditions $(\star,\lozenge)$,
$b\equiv-1/3\epsilon_0L^3$ for all $T$.  All curves end at
$r=2^{1/6}\,a$, where the particles interact with the truncated
Lennard-Jones potential.}
\end{figure}

\subsubsection{Pair potential}
\label{pot-subsec}
We begin the analysis of the algorithm by verifying that the effective
interaction between particles is in agreement with Eq. (\ref{ew-eq}).
We evaluated the pair potential by placing two particles of opposite
charge in the simulation cell and measuring the distribution $P(r)$ of
separations between the particles for different temperatures and
system sizes. The negative logarithm of $P(r)$, which is proportional
to the pair potential, is shown in Fig.~\ref{pot-fig} together with
the curve Eq.~(\ref{ew-eq}). To include self-avoidance, a purely
repulsive truncated Lennard-Jones potential of range $a$ was added to
the interaction. Overall, we find excellent agreement between
Eq.~(\ref{ew-eq}) and the numerical curves. For Maxwell boundary
conditions, the quadratic correction depends on temperature, and the
two temperatures shown in Fig.~\ref{pot-fig} correspond to the
limiting values of Eq.~(\ref{ew-eq}).

The Ewald convention with `tinfoil'' boundary conditions has also been
simulated by performing additional integration over the mean field
$\bar{ \bf E}$. In this case, all curves collapse onto the $l_b <<L$
Maxwell result for all temperatures.  To reduce lattice artifacts at
short distances, a dynamic subtraction (that will be discussed in a
subsequent section) was employed for the curves of Fig.~\ref{pot-fig}.

\subsubsection{Screening}
\begin{figure}
\includegraphics[width=8cm]{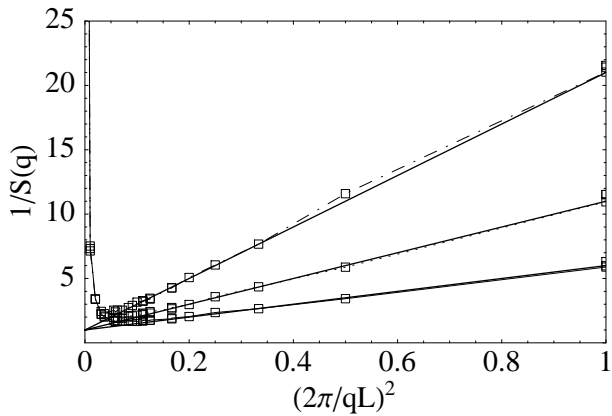}
\caption{\label{debye-fig}Static charge-charge structure factor for a
simple Coulomb gas at $T=1.25 T^\star$ and three different densities
$\rho =0.05\,a^{-3},$$\rho= 0.1\,a^{-3}$, and $\rho= 0.2\,a^{-3}$. The
slopes at large values of $1/q^2$ increase linearly with density as
expected for Debye screening.}.
\end{figure}

One of the characteristic features of systems containing mobile
charges is the phenomenon of screening; the effective interactions
decay exponentially with the inverse Debye screening length
$\kappa^{2}=ne^2/\epsilon_0k_BT$ \cite{chaikin}, where $n$ denotes the
density of the charge carriers.  At high enough temperatures the
static charge-charge structure factor of a screened system has the
form
\begin{equation}
S(q)= \frac{e^2q^2}{\kappa^2+q^2}.
\label{debye-eq}
\end{equation}
In order to test whether our algorithm reproduces correctly this
essential property of charged systems, we consider a globally neutral
polyelectrolyte composed of positive and negative particles (together
with a short ranged Lennard-Jones potential) and equilibrate it at the
temperature $T=1.25T^\star$. We then measure the static structure
factor
\begin{equation}
S(q)= \langle s({\bf q})s(-{\bf q})\rangle,\,\,  s({\bf q})=\frac{1}{\sqrt{N}}\sum_i e_i\exp{(i{\bf r}_i{\bf q})}
\end{equation}
and compare its form to Eq.~(\ref{debye-eq}) in Fig.~\ref{debye-fig}
by plotting $1/S(q)$ as a function of $1/q^2$. As can be seen, the
long wavelength data falls along a straight line, whose slope varies
with particle density. The solid lines have slope $\kappa^2L^2/4\pi^2$
and are in good agreement with the numerics.

\subsection{Efficiency}
Having shown that the algorithm behaves correctly, we now turn to a
study of the dynamics. As already discussed in \cite{acm2}, the
Monte-Carlo update rules lead to a diffusive dynamics of the
longitudinal and transverse fields, and we wish to explore how quickly
our off-lattice algorithm equilibrates various modes of the system.

\subsubsection{Dynamics of free charges}
\begin{figure}
\includegraphics[width=8cm]{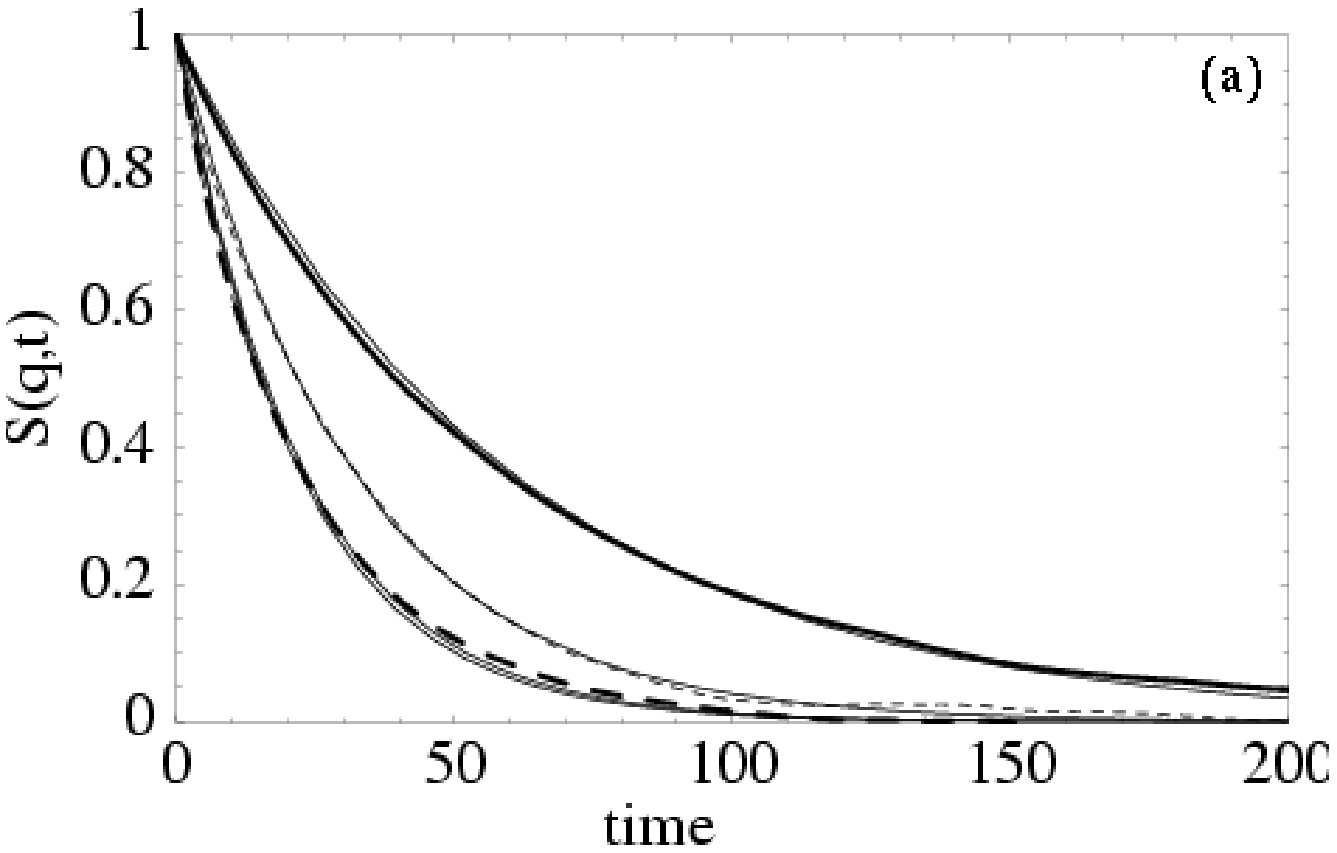}
\includegraphics[width=8cm]{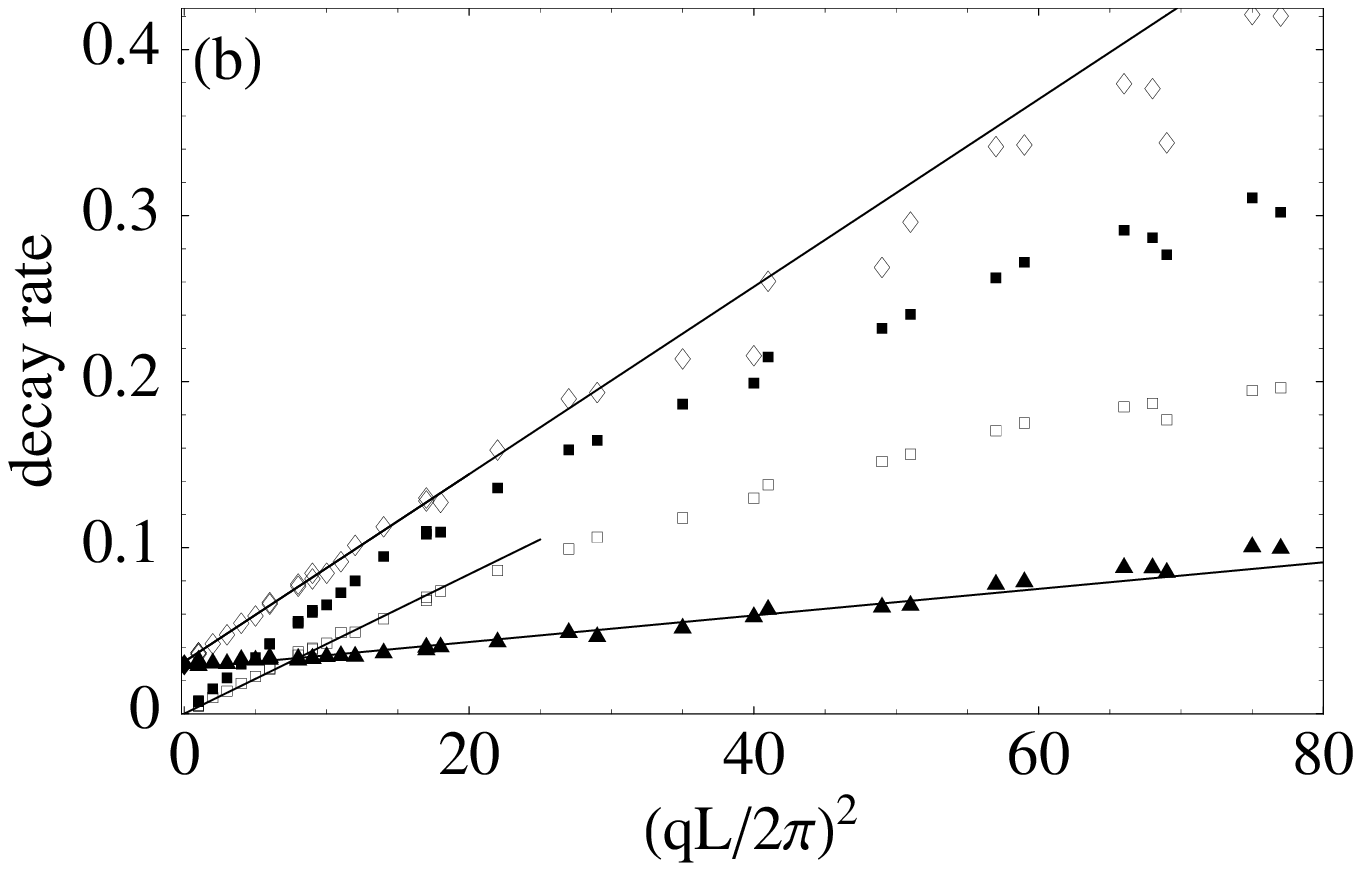}
\caption{\label{disp-fig}(a) Relaxation of particle (thick solid
line), charge (thin dashed line) and transverse field (thick dashed
line) correlations as well as exponential fits.  System size
$L=20\,a$, $\rho=0.2\,a^{-3}$, $T=1.25T^\star$, mode (1,1,1). (b)
Dispersion relations for particle $(\square)$, charge
$(\blacktriangle)$ and transverse field $(\lozenge)$. Also shown is
the density dispersion relation for a neutral system $(\blacksquare)$
with otherwise identical parameters. }
\end{figure}

We investigate the dynamics of the simple Coulomb gas shown in
Fig.~\ref{debye-fig} by studying the time decay of correlations in
Fourier space. We measure the dynamical structure factors $S({\bf
q},t)=\langle s({\bf q},t)s(-{\bf q},0)\rangle$ with $s({\bf
q},t)=\frac{1}{\sqrt{N}}\sum_i e_i\exp{(i {\bf q}\cdot {\bf r}_i(t))
}$, where the weights $e_i$ are replaced by unity when measuring
density correlations.  We also examine correlations in the transverse
electric field (that part of the field for which the discrete
divergence, Eq. (\ref{intgauss-eq}) is zero).  The correlation
functions for a single wavevector are shown in
Fig.~\ref{disp-fig}(a). All modes decay exponentially; time is
measured in ``particle sweeps''; in one time unit we attempt to update
each particle once together with a variable number of plaquettes (see
below).

By fitting the correlation functions to exponentials, we extract decay
rates for selected modes and plot the resulting dispersion relations
in Fig.~\ref{disp-fig}(b). As can be seen, the density modes
$(\square)$ are diffusive at small wavevectors, $\omega\propto q^2$,
but the longitudinal $(\blacktriangle)$ and transverse $(\lozenge)$
modes of the electric field have a gap at $q=0$, $\omega\propto
\omega_0+ q^2$.  This behavior can be compared to the continuum
equations describing a gas of free charge carriers (see also
ref.~\onlinecite{acm2}). The mean current is driven by the electric field
and the charge density gradient,
\begin{equation}
{\bf J}=ne^2\alpha{\bf E}-D\nabla \rho.
\end{equation}
$n$ is the number density, 
$D$  a diffusion coefficient that is related to the mobility
$\alpha$ via the Einstein relation $D=k_BT\alpha$. We insert this
relation into the diffusion equation (\ref{diff-eq}) to study the
dispersions of the longitudinal and transverse field components.

First, we take the curl of Eq.~(\ref{diff-eq}) and Fourier transform
to find the relation
\begin{equation}
(i\omega+\epsilon_0{\bf q}^2/\xi+ne^2\alpha/\epsilon_0){\bf \hat E}_{tr}=0.
\end{equation}
As observed in Fig.~\ref{disp-fig}(b), the transverse modes diffuse
with a diffusion constant that is determined by $\xi$, a
characteristic relaxation time scale related to the frequency of
updates of the transverse field components. In the absence of charges,
the relaxation time $\omega\rightarrow 0$ as $q\rightarrow 0$, but the
presence of a charge density induces a gap
$\omega_0=Dne^2/k_BT=D\kappa^2$ in the spectrum. The origin of this
gap is the screening of the interactions on a scale given by the
inverse Debye screening length $\kappa^2$.

For the longitudinal field components, we find that
\begin{equation}
(i\omega+D{\bf q}^2+ne^2\alpha/\epsilon_0){\bf q}\cdot{\bf E}=0.
\end{equation}
Again we find a gap in the spectrum with the same frequency $\omega_0$
as for the transverse modes.  We have also computed numerically the
relaxation of the field correlations $\langle \bar{\bf E}(t)\bar{\bf
E}(0)\rangle$ in real space, which corresponds to the $q=0$ mode in
Fig.~\ref{disp-fig}(b). This frequency coincides with $\omega_0$.

Fig.~\ref{disp-fig} provides the basic demonstration of the efficiency
of the algorithm. The slowest relaxation time of the system
corresponds to particle diffusion and scales quadratically with system
size; any meaningful statistical average must equilibrate the sytem
over that timescale.  The charge and transverse fluctuations relax
much faster than the density fluctuations, and in Monte-Carlo one is
free to tune the ``diffusion constant'' (the slope of the straight
line) for the transverse field by adjusting the relative frequency of
plaquette to particle updates. If we update the transverse field very
often then we converge to the classic algorithms with instantaneous
propation of the electric field. However, even very rare transverse
updates will still equilibrate the transverse field components.

For Fig.~\ref{disp-fig}, we have chosen a high frequency of transverse
mode updates, where we update typically 2/3 of all plaquettes in one
particle sweep. Nevertheless, we estimated only about 15\% CPU time is
needed for this part of the simulation, all other time was consumed by
the particle updates (essentially charge interpolation plus the
Hamiltonian path).  We dropped the frequency of plaquette updates by a
factor of 10 and observed no change in the relaxation of the density
and charge correlations. This result indicates that in a reasonably
dense system, the particle motion alone is already performing a
substantial part of the integration over the transverse modes.  Our
simple analytic theory predicts exactly this: even in the limit of
slow propagation of the transverse field the longitudinal dynamics
equilibrate.

As an additional illustration of the efficiency of the algorithm, we
replaced the charged particles of Fig.~\ref{disp-fig}(b) with neutral
particles and repeated the simulation with otherwise identical
parameters. The particle dispersion $(\blacksquare)$ shows that the
neutral particles diffuse about two times faster, since they are not
slowed down by the electric field. We conclude that the price for
simulating this dense charged system is an overhead of only about a
factor of 2 in particle sweeps (though clearly in a neutral system one
does not need charge interpolation which is relatively costly in cpu
time).

\begin{figure}
\includegraphics[width=8cm]{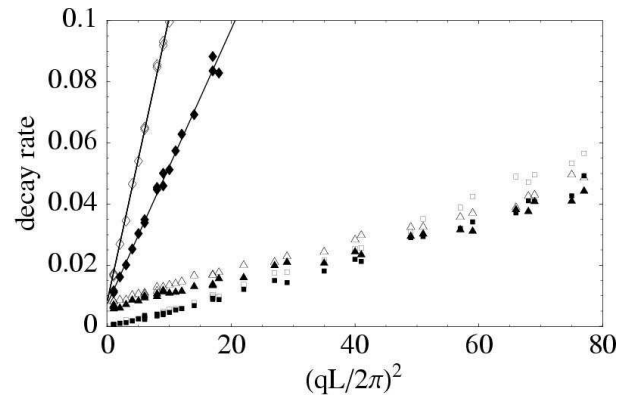}
\caption{\label{dilute-fig}Dispersion relations in a simulation with
system size $L=20\,a$, $\rho=0.01\,a^{-3}$, $T=0.125T^\star$ for
particles $(\square)$, longitudinal $(\triangle)$ and transverse
fields $(\lozenge)$. For the solid symbols the transverse fields were
updated less frequently (see text).}
\end{figure}

It is clear that the algorithm exhibits strict ${\cal O}(N)$ scaling
(at least away from critical points), since the work for moving a
particle is purely local and consists essentially in the charge
interpolation. One might object that the algorithm is surely
inefficient at very low densities: the effort needed to update the
plaquettes must become overwhelming. To characterize the speed of our
algorithm in such an unfavorable situation, we contrast the high
temperature, high density simulation of Fig.~\ref{disp-fig} with a
simulation at a lower temperature $T=0.125T^\star$ and density
$\rho=0.01\,a^{-3}$ in Fig.~\ref{dilute-fig}. It is now necessary to
perform more plaquette updates than before; most particles have
condensed into pairs that move slowly and cannot easily integrate over
the transverse field. For the open symbols in Fig.~\ref{dilute-fig} we
update each plaquette once per particle sweep and achieve a very high
relaxation rate for the transverse field. The computational effort for
the transverse updates is about 70\% of the total runtime in this
case. We dropped the frequency of transverse updates by a factor of
two (filled symbols), but the longitudinal and transverse relaxation
rates only dropped by about 10\%.

We should note that we are using a very simple Monte-Carlo
procedure in this low density system, which chooses plaquettes
randomly in the simulation cell.  One can easily bias the choice of
the plaquette to be in the neighborhood of a particle without
violating detailed balance. Initial tests on our lattice
implementation \cite{acm} have shown substantial increase in
efficiency at very low densities (volume fraction of $10^{-4}$) with
such a choice.  We therefore anticipate that simple, local
optimizations will greatly improve the performance of the algorithm
for very dilute systems at low $T$, but leave a detailed exploration
of this point to future work.

\subsubsection{Dynamics of polar molecules}
\begin{figure}
\includegraphics[width=8cm]{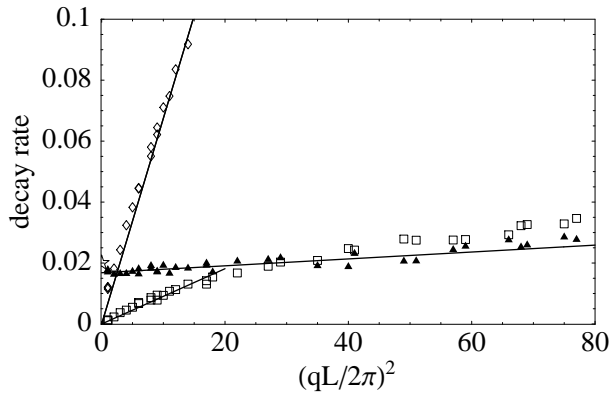}
\caption{\label{dip-fig}Dispersion relations for particle $(\square)$,
charge $(\blacktriangle)$ and transverse field $(\lozenge)$ for simple
dipoles. System size $L=20\,a$, $\rho=0.2\,a^{-3}$, $T=1.25T^\star$,
$k=8\pi T^\star$.}
\end{figure}

In the previous paragraph, we argued that relaxation of the simple
Coulomb gas is aided by screening due to the gap in the relaxation
spectrum. It is therefore interesting to ask how the algorithm
performs on a non-screening system. To this end, we create polar
molecules by coupling two particles with opposite charge with a
harmonic spring of the form $V_b(r)=k(r-r_0)^2$.  In
Fig.~\ref{dip-fig}, we show the dispersion relations for this polar
system.

Comparing with Fig.~\ref{disp-fig}, we observe that the spectrum for
the transverse modes $(\lozenge)$ no longer has a gap, but rather
extrapolates to zero at $q=0$. However, the dispersion continues to be
diffusive, and the transverse modes can be relaxed at any desired rate
by varying the number of plaquette updates per sweep. The longitudinal
$(\blacktriangle)$ spectrum continues to exhibit a gap, but its
magnitude is now determined by the spring stiffness governing the
internal vibrations of the molecule.  The efficiency of the algorithm
is independent of the presence of screening, and the relaxation of the
transverse modes is not impeded by the longitudinal field.

\subsection{Numerical stability}
\begin{figure}
\includegraphics[width=8cm]{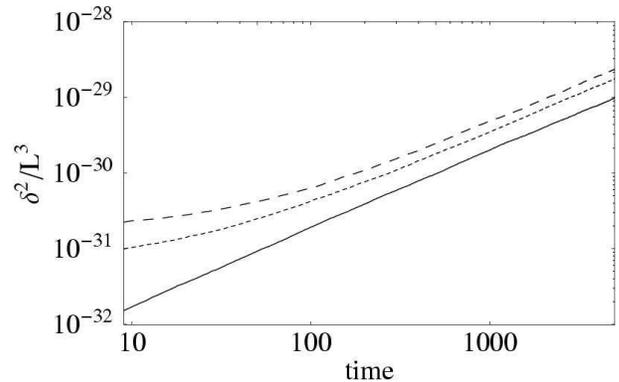}
\caption{\label{stab-fig}Squared deviation from Gauss' law/site as a
function of time in a simulation of size $L=20\,a$, where the number of
particles $N=0$ (solid), $N=800$ (dashed) and $N=1600$ (long
dashed). }
\end{figure}

Another appealing feature of the algorithm besides its simplicy is its
numerical stability. Every upate of the field on a link, be it through
particle motion or plaquette updates, involves a numerical error of
order the machine precision (typically ${\cal O}(10^{-16})$). Although
this error is cumulative over time, it increases so slowly that the
accuracy should remain acceptable even for the longest run times. To
investigate the stability of our code, we averaged the deviation of
the field configuration from Gauss's law Eq.~(\ref{intgauss-eq}),
$\delta^2=(a^2\sum_j E_{i,j}-e_i/\epsilon_0)^2$ over the entire
simulation cell and plot this quantity as a function of time in
Fig.~\ref{stab-fig}. Time was again measured in Monte-Carlo sweeps. We
find a diffusive growth for the error/site, $\delta^2/L^3={\rm
const.}\times 10^{-32}t$. The constant is smallest for an empty system
and increases slowly with the number of particles, but is always less
than one. Since the equilibration time is of order ${\cal O}(L^2)$,
even the largest systems  equilibrate with acceptable errors.

\section{Dynamic subtraction of lattice artefacts}
\label{yuk-sec}
Many conventional fast electrostatic methods (and the present
algorithm) interpolate the point charges onto an electric grid in
order to exploit the fast Fourier transform (FFT) \cite{holm2} or
multigrid algorithms \cite{darden2}.  Such an interpolation introduces
a set of artefacts that have to be controlled: at short scales, the
interaction between the charge clouds will in general differ from that
of point charges; furthermore, the self energy ${\cal U}_{self}$ of an
interpolated charge distribution $\rho_{int}({\bf r})$ becomes a
function of its position with respect to the mesh. In general,
\begin{eqnarray}
\nonumber
{\cal U}_{self} &=&{1\over 2}\int d^3r d^3r'G({\bf r}-{\bf r}')
\rho_{int}({\bf r})\rho_{int}({\bf r}')\\
&=& {1\over 2}\int {d^3 q \over (2 \pi)^3} \hat{G}({\bf q}) S_{int}({\bf q})
\label{self-eq}
\end{eqnarray}
where $S_{int}({\bf q})$
is the structure factor of the interpolation of the charge to the
mesh. For the Coulomb interaction, the Green functions are $G_{\rm
coul}(r)\sim 1/r$ and $\hat{G}_{\rm coul}({q})\sim 1/{q^2}$ in real
and Fourier space, respectively, so that the integral diverges
linearly for $q\rightarrow \infty$.
This leads to a periodic potential which has an amplitude $\Delta U_s$
of order ${\cal O}(e^2/\epsilon_0 a)$ for mesh size $a$. Use of a
finer mesh leads to {\sl worse} results with the simplest
interpolation schemes.  The periodic potential tends to ``trap'' the
particles in the center of the grid cubes where the self-energy is
lowest.

The conventional remedy to the situation consists simply in spreading
out the charge over a wider range of sites, thereby reducing the
self-energy artifacts. This is typically done with a convolution step
that distributes the charges over several hundred  sites using
Gaussians \cite{darden2} after interpolation to the lattice.  These
convolutions are easy to perform in Fourier space, but are complicated
and time consuming in real space. They would also drastically reduce
the speed of our algorithm, since the Hamiltonian path would have to
visit many more sites in order to properly update the field
configuration.

Here, we introduce a dynamic correction that leads to an effective
convolution of the charge distribution with minimal computational
overhead. For simplicity, we initially work in a continuum picture
before discretizing and consider the unconstrained scalar functional
\begin{equation} {\cal F}_Y [\psi]=
\int \left[ {\epsilon_0 \over 2}[ (\nabla \psi)^2 + \mu^2\psi^2] -
\rho \psi \right] d^3 r\label{yuk-eq}
\end{equation}
Minimizing this functional with respect to the field $\psi$ leads to
the Yukawa equation
\begin{equation}
(\nabla^2 -\mu^2)\psi_Y = -\rho/\epsilon_0.
\end{equation}
Substituting the solution $\psi_Y$ into Eq.~(\ref{yuk-eq}), we find
a {\em negative} Yukawa Green function.
The negative sign of the Green function is a feature of the scalar energy
functional that we exploit here. By coupling the charge density $\rho$
in our simulation to the scalar functional Eq.~(\ref{yuk-eq}), we
generate the partition function
\begin{equation}
{\cal Z}_{Y}({\bf r})= \int {\cal D}\psi e^{-{\cal F}_Y[\psi]({\bf r})/k_BT}
\end{equation}
and the particles interact with the Yukawa potential
\begin{equation}
\label{yukpot-eq}
V_Y(r) = - {e_1 e_2 \over 4 \pi \epsilon_0 r} e^{-\mu r}.
\end{equation}

The essence of our dynamic subtraction algorithm is to couple the
charges {\em simultaneously} to the unconstrained functional
Eq.~(\ref{yuk-eq}) as well as the constrained vector functional,
Eq.~(\ref{func-eq}). The total partition function then reads
\begin{equation}
{\cal Z}({\bf r})={\cal Z}_{Coulomb}({\bf r})\times {\cal Z}_{Y}({\bf r})\times {\rm const.}
\end{equation}
and implies an effective interaction between two charges $e_1$ and
$e_2$ given by
\begin{equation}
\label{effpot-eq} 
V(r) =  {e_1 e_2 \over 4 \pi \epsilon_0 r} (1-e^{-\mu r}).
\end{equation}
At large separations we find the Coulomb interaction; at small
distances the potential has been regularized. The corresponding
Green's function for eq.(\ref{effpot-eq}) is the sum of the Green's
functions for Coulomb and Yukawa interaction,
\begin{equation}
\label{scheme1-eq}
\hat{G}({\bf q})=\frac{1}{{\bf q}^2}-\frac{1}{{\bf
q}^2+\mu^2}=\frac{\mu^2}{{\bf q}^2({\bf q}^2+\mu^2)}
\end{equation}
By inserting eq. (\ref{scheme1-eq}) into Eq.~(\ref{self-eq}), we find
that the self-energy is now finite for $a \rightarrow 0$.  We can
interpret the factor that multiplies the bare Coulomb Green function
on the right of eq.(\ref{scheme1-eq}) as a structure factor
$S_{conv}({\bf q})= \mu^2/(q^2+\,u^2)$ describing a convolution of the
original charge density. In real space, this convolution function is
\begin{equation}
\int_{-\infty}^{\infty} {d^3 {\bf q}\over (2 \pi)^3} S_{conv}({\bf
q})^{1/2} {e^{-i {\bf q \cdot r}}}=\frac{\mu^2}{2\pi^2}\frac{K_1(\mu
r)}{r},
\end{equation}
where $K_1$ is a modified Bessel function. 

Use of the scheme Eq.~(\ref{effpot-eq}) allows us to construct an
efficient algorithm, since it requires only the introduction of a
scalar field with Monte-Carlo moves and coupling to the charge density
according to Eq.~(\ref{yuk-eq}). The additional overhead of the
dynamics of this supplementary field is small, since we reuse the same
values of the interpolated charges for the vector and scalar fields.
The run time complexity of the dynamic correction is independent of
$\mu$. Small values of $\mu$ give maximum smoothing, but also strongly
alter the interaction at small distances. In general, one therefore
wants to remove the Yukawa interaction on a scale of $1/\mu$ by adding
$-V_Y(r)$ to the interaction. This can be conveniently done together
with all the other short range (e.~g. van der Waals) interactions of
the system. The choice of $\mu$ therefore gives us a freely tunable
tradeoff between precision and efficiency.
\begin{figure}
\includegraphics[width=8cm]{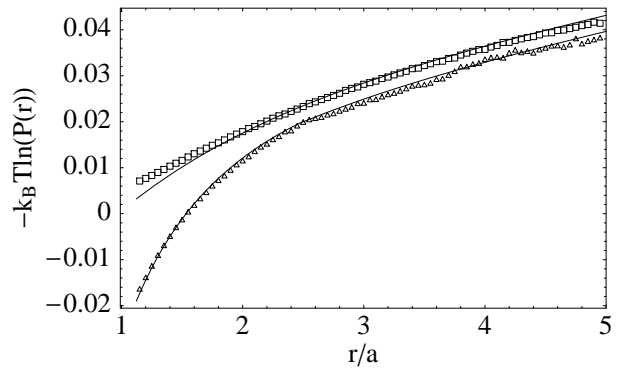}
\caption{\label{yukpot-fig}Pair potential for combined Coulomb/Yukawa
interaction at $T=0.0625T^\star$, $L=10\,a$, $\mu=0.75a^{-1}$. The top
curve $(\square)$ shows the full combination in agreement with
$V(r)={\rm const} -(1-\exp[-\mu r])/4\pi \epsilon_0 r-
r^2/3\epsilon_0L^3)$ (solid line). In the lower curve $(\triangle)$,
the Yukawa interaction was removed for distances up to $r\le
2.5a$. For the curves shown in Fig.~\ref{pot-fig} the Yukawa
interaction was removed up to $L/2$.}
\end{figure} 
This algorithm reminds us of early attempts at creating finite field
theories in the 1940's \cite{pais} using different particles to cancel
divergences in physical quantities. Here it is the ground state
energies of two Gaussian field theories (but with scalar and vector
symmetries) which have compensating divergences in the continuum
limit, leading to a finite theory when they are combined.

We have implemented the above method in our off-lattice Monte-Carlo
simulation to test its validity and efficiency. To this end,
Monte-Carlo moves for the second auxiliary field $\psi$ were
introduced to generate the effective interaction
Eq.~(\ref{effpot-eq}). We first verified this interaction by
extracting the pair potential from the distribution of separation
$P(r)$ of a pair of charges as in Sec.~\ref{pot-subsec}. The top curve
in Fig.~\ref{yukpot-fig} shows the combined Yukawa and Coulomb
interaction for a value of $\mu=1$ in the low temperature limit of
Maxwell boundary conditions. The Yukawa field was updated much more
rarely than the transverse electric field since the massive field
relaxes rapidly. The curve agrees well with the analytic expression
$V(r)={\rm const} -(1-\exp[-\mu r])/4\pi\epsilon_0 r-
r^2/3\epsilon_0L^3)$ except at small distances where the interpolated
charges overlap.  The Yukawa potential can be subtracted
Eq.~(\ref{yukpot-eq}) explicitly at short distances. The lower curve
shows a case where this was done for $r<2.5 a$; the numerical points
agree well with the Coulomb potential.  Note that this correction at
small scales should in fact not be neccessary in realistic
applications that use a grid size $a$ that is much smaller than the LJ
particle radii.

\begin{figure}
\includegraphics[width=8cm]{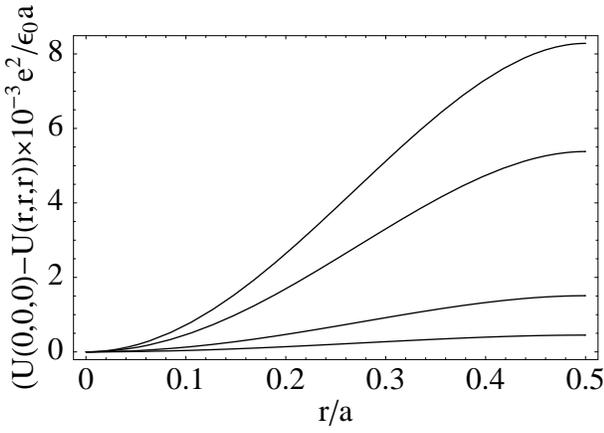}
\caption{\label{trappot-fig}Self energy along the diagonal of a grid
cube from corner $(0,0,0)$ to center $(0.5,0.5,0.5)$ for the pure
Coulomb interaction (lowest curve) and the combined Coulomb/Yukawa
interaction for $\mu=3a^{-1}$, $\mu=1a^{-1}$, and $\mu=0.5a^{-1}$
(highest curve). Charges were interpolated using B-splines, and the
energies were calculated using the lattice Green function
Eq.~(\ref{latgreen-eq}).}
\end{figure}

We now have to show that the imposition of an additional Yukawa
interaction  also reduces the self-energy artifacts. The
most dramatic manifestation of these artifacts occurs through a
trapping of particles in the cell center due to the (periodic) energy
barrier $\Delta U_s=U_{\rm corner}-U_{\rm center}$, where $U_{\rm
center}$ and $U_{\rm corner}$ are the self energies of a particle
located at the center or corner of a grid cube. The self energy of a
particle whose charge is interpolated onto the cubic grid with the
B-spline interpolation Eq.~(\ref{chargeint-eq}) can be calculated
using the lattice Green function for the Yukawa interaction,
\begin{eqnarray}
\label{latgreen-eq}
G_{\mu}({\bf r}) &=& \int {d^3 {\bf q}\over (2 \pi)^3}  {e^{i {\bf q \cdot r}} \over
6 - 2 \cos {q_x} -  2 \cos {q_y} -  2 \cos {q_z} + \mu^2} 
\end{eqnarray}
We evaluated Eq.~(\ref{latgreen-eq}) numerically using
Mathematica. Fig.~\ref{trappot-fig} shows the self-energy along the
diagonal of a grid cube for for the combined Coulomb/Yukawa
interaction for several values of $\mu$. The amplitude of this
potential is largest for the pure Coulomb case and is drastically
reduced when $\mu=0.5a^{-1}$

\begin{figure}
\includegraphics[width=8cm]{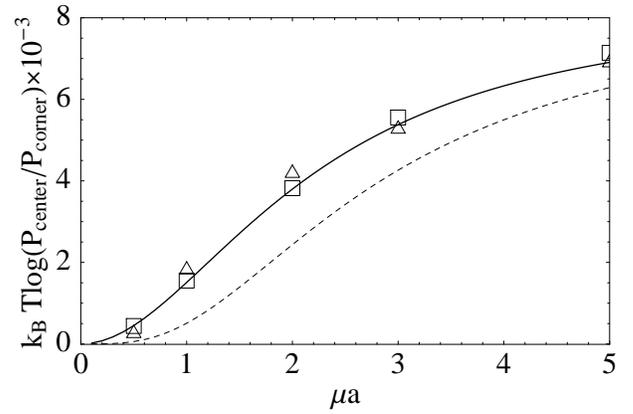}
\caption{\label{trapping-fig}Reduction of the self energy barrier
leading to trapping of particles in cell centers. Numerical points
correspond to temperatures $T=0.0625T^\star\, (\square)$ and $T=0.125
T^\star\, (\triangle) $.  The probabilities were measured by counting
particles in cubic regions centered in center and corner of size
$0.4^3a^3=0.064a^3$. The lines show the expected reduction of the
energy barrier based on a semi-analytic evaluation of the self-energy
using the scheme Eq.~(\ref{effpot-eq}) (solid line) and
Eq.~(\ref{best-eq}) (dashed line).}
\end{figure}

In our Monte-Carlo simulations, $\Delta U_s$ has been  extracted
from the ratio of probabilities to find the particle in the center or
corner, respectively. In Fig.~\ref{trapping-fig}, we show the
effective self-energy barriers obtained in this way as a function of
$\mu$. As expected from Fig.~\ref{trappot-fig}, the barrier drops
rapidly with decreasing $\mu$, and almost complete elimination is
achieved at a value of $\mu=0.5a^{-1}$. The solid line in
Fig.~\ref{trapping-fig} was obtained by computing the self energy
difference of particles located in the center and corner of the grid
cube as shown in Fig.~\ref{trappot-fig}.  The numerical points agree
well with this result.
 
From the numerical integration of Eq.~(\ref{latgreen-eq}), we find
that the amplitude of the periodic energy barrier varies as $\Delta
U_s(\mu)=2\times 10^{-3}(\mu a)^2T^\star$ for small $\mu$.
Even better smoothing should be possible if one uses a mixture of Green
functions that decays even faster for large wavevectors than
Eq.~(\ref{scheme1-eq}): Let us consider
\begin{equation}
\hat{G}({\bf q}) = {2 \mu^4 \over {\bf q}^2 ({\bf q}^2 + \mu^2) ( {\bf
q}^2 + 2 \mu^2)} \label{effective},
\end{equation}
 The corresponding real space potential is 
\begin{equation}
V(r) = {e_1 e_2 \over 4 \pi \epsilon_0 r} ( 1 - 2 e^{-\mu r} + e^{-
\sqrt{2} \mu r}),
\label{best-eq}
\end{equation}
which can be generated by coupling the charges to three fields,
respectively a constrained vector, a scalar and an {\sl unconstrained} vector.
 This unconstrained vector functional reads
\begin{equation}
{\mathcal F}_{\mu}[{\bf E}]=\int \left[\frac{\epsilon_0{\bf E}({\bf
r})^2}{2}-\frac{(\epsilon_0{\rm div\,}{\bf E}({\bf r})-\rho({\bf
r}))^2}{2\mu^2}\right]d^3r.
\label{func-eq2}
\end{equation}
which leads to a Yukawa Green function with a positive prefactor.  The
dashed line in Fig.~\ref{trapping-fig} shows the energy barrier as a
function of $\mu$ for this scheme.  The inhomogeneity in the self
energy is smaller than when using only the scalar Yukawa functional,
and the barrier now varies as $\Delta U_s(\mu)=1.5\times 10^{-3}(\mu
a)^4T^\star$ for small $\mu$. Although we have not implemented this
full scheme in our Monte-Carlo simulation, it appears to be the
strategy of choice when optimizing the algorithm for truly atomistic,
high accuracy simulations.  We note that the sucess of the method
requires that both the vector and scalar functionals implement
equivalent discretizations of the Laplacian operator. In our code this
is the simplest 7-point difference scheme.

We would like to emphasize that the present strategy is general and
can also be used in combination with other Coulomb algorithms. Methods
relying on solutions of the Poisson equation could simulate an
additional Yukawa equation and add the resulting potentials or
forces. This seems to be rather attractive, since the charge spreading
step amounts to a significant portion of the total computational
effort in multigrid formulations \cite{darden}.

\section{Conclusions}
We have implemented a Monte-Carlo algorithm for charged
particles moving in the continuum. Instead of minimizing the
electrostatic energy, the algorithm uses particle moves in
combination with constrained auxiliary field updates whose dynamics
are inspired by the full Maxwell equations. Unlike 
in electrodynamics, however, the field propagates diffusively. We showed
that the algorithm efficiently equilibrates charged systems such as
electrolytes and polar fluids. It is most efficient at higher
densities and temperatures, where the particle motion alone can
perform a substantial part of the integration over the transverse
modes of the electric field. Even at lower temperatures, the cost for
the transverse integration remained acceptable.

We also introduced an approach to treat lattice artifacts that
arise from the charge interpolation onto a grid. By introducing an
additional negative Yukawa interaction, the periodic potential arising
from the self-energy of the particles can be reduced to any desired
accuracy. The scheme is completely general, and other Coulomb
algorithms could also benefit from the method.

The principal advantages of the present approach are
\begin{itemize}
\item True ${\cal O}(N)$ scaling with the number of particles in the
system. No such algorithm is available for Monte-Carlo simulations at
present. Many problems involving electrostatics that could previously
only be treated efficiently with molecular dynamics can now be treated with
Monte-Carlo, which is often easier to program and faster due to larger
time steps.

\item Purely local computations are required to advance particle and
field degrees of freedom, making the algorithm an ideal candidate for
parallelization on large compute clusters or supercomputers. 
\end{itemize}

In the present form, the algorithm can treat coarse-grained models of
e.~g. polyelectrolytes or membranes. It now has to pass more stringent
tests with realistic model systems and truly atomistic simulations.

\section{Acknowledgments}
We thank R.~Everaers and L.~Levrel for very stimulating discussions of
this work. JR thanks the CNRS for financial support through a ``poste
rouge'' fellowship.

\appendix
\section{Boundary conditions for charged periodic systems}
We first summarize the boundary conditions employed for Ewald
 summation and then discuss their relation to the Maxwell boundary
 conditions, which are the natural choice for our algorithm. The Ewald
 formula for the electrostatic energy of periodic a system is
\begin{equation}
U= \frac{1}{8\pi\epsilon_0}\sum_{i,j\neq i}e_ie_j\psi_{per}({\bf
r}_i-{\bf r}_j)+U_{self}+U_{dip}
\end{equation} 
The self-energy, $U_{self}=\kappa/\sqrt{\pi}\sum_ie_i^2$, is constant
and therefore irrelevant for Monte-Carlo dynamics.  The periodic Ewald
potential $\psi_{per}$ consists of sums in real and Fourier space:
\begin{eqnarray}
\nonumber
\psi_{per}={4\pi\over L^3}\sum_{{\bf G \ne 0}} {
{\exp (-G^2/4\kappa^2 + i {\bf G \cdot r}})
\over {G^2}
}\\
-{\pi \over \kappa^2 V}
+\sum_{\bf R} {{\rm erfc}(\kappa |{\bf r + R}  |)
\over
|{\bf r + R}|}
\end{eqnarray}
The result of the sum is independent of the range parameter, $\kappa$.
At short distances, this expression can be expanded \cite{fraser}, giving
\begin{equation}
\psi_{per}  = const + {1\over r} + {2\pi r^2\over 3L^3}+{\cal O}(r^4).
\label{short-eq}
\end{equation}
The second term is of the same order in $1/L$ as the dipole energy
\begin{equation}
\label{dipcor-eq}
U_{dip}={1 \over 2\epsilon_0(2 \epsilon' +1)L^3} \left| \sum_i e_i
{\bf r}_i \right|^2.
\end{equation}
This energy depends on the dielectric constant $\epsilon'$ of the
surrounding medium. Two common choices for the dielectric constant are
$\epsilon^\prime=1$ (``vacuum'') and $\epsilon^\prime=\infty$
(``tinfoil'').

A term comparable to $U_{dip}$ arises when using Maxwell boundary
conditions: In Sec.~\ref{alg-sec}, we saw that the field dynamics obey
a Maxwell equation,
\begin{equation}
{\partial {\bf E} \over \partial t}= 
-{\bf J}/\epsilon_0 + {\rm curl}\, {\bf B}/\mu_0.
\end{equation}
Integrating this equation over space and time, we find the relation
between the ${\bf q}=0$ component of the electric field and the dipole
moment, ${\bf d}$ 
\begin{equation}
\bar{\bf E}(t)=-\frac{1}{L^3}\int dt\int d^3r\, {\bf
J}/\epsilon_0=-{\bf d}/L^3\epsilon_0,
\end{equation}
since the integral of the curl of a periodic function is zero.  The
contribution of $\bar {\bf E}$ to the electrostatic energy is
\begin{equation}
{\cal U}_d = {\epsilon_0 \over 2} \int \bar{\bf E}^2 = 
 {{\bf d}^2 \over 2L^3\epsilon_0}, 
\end{equation}
The longitudinal components of $\bf E$ for ${\bf q} \ne 0$ are given 
by the gradient of $\psi_{per}/4\pi \epsilon_0$.

Note that there is an important difference between the conventional
dipole moment of the simulation cell, $\sum e_i {\bf r}_i$ and ${\bf
d}$: A charge can wind about the periodic system giving equivalent
configurations with different energies.  These two definitions of the
dipole moment are only the same modulo a Bravais lattice vector
\cite{caillol}:
\begin{equation}
{\bf d} = \sum_i e_i {\bf r}_i + |e| L \{l,m,n\}.
\end{equation}
for arbitrary integers $l$, $m$, $n$.  

Consider now a pair of particles with opposite charges separated by
${\bf r}$, simulated by our algorithm.  If we sum over all equivalent
configurations we find the relative statistical weight of a given
state
\begin{equation}
{\cal Z}_0 = \sum_{l,m,n} e^{-e^2({\bf r}+ L\{l,m,n\})^2/2k_BT\epsilon_0 L^3},
\label{sum}
\end{equation}
At low temperatures and with ${\bf r}$ small the sum Eq.~(\ref{sum})
is dominated by a single minimum energy term with $l=m=n=0$, thus
\begin{equation}
{\cal Z}_0= e^{ -e^2{\bf r}^2/2k_BT\epsilon_0 L^3}\quad {\rm
for}\quad  k_BT\ll e^2 /4 \epsilon_0 L
\end{equation}
The corresponding free energy is
\begin{equation}
\label{lowt-eq}
F = - kT \ln {\cal Z}_0 = e^2{\bf r}^2/2\epsilon_0 L^3.
\end{equation}

We now combine the low temperature result with the short distance
expansion of the Ewald potential for two particles by adding
Eq.~(\ref{lowt-eq}) to Eq.~(\ref{short-eq}). We find
\begin{equation}
V(r)= -e^2\left({1 \over 4\pi\epsilon_0 r} + 
{r^2\over 6\epsilon_0 L^3} \right)+{e^2 r^2 \over 2\epsilon_0 L^3}.
\end{equation}
We thus expect a net quadratic correction
$e^2r^2/3\epsilon_0 L^3$ to the $1/r$ interaction as in
Eq.~(\ref{ew-eq}).

In the opposite limit of high temperatures, the charges are mobile and
the current can wind around the simulation cell many times. In this
limit
\begin{equation}
{\cal Z}_0\sim {\rm const.}\quad {\rm
for}\quad k_BT\gg e^2 /4 \epsilon_0 L.
\end{equation}
There are no dipole contributions in this limit, and the potential is
that of Eq.~(\ref{short-eq}).

Note that the above arguments hold for free charges that allow the
current to flow freely; for bound charges or dipoles the current does
not lead to multiple inequivalent states for the field $\bar {\bf E}$.
It is possible to impose the tinfoil boundary condition in our
algorithm by introducing a third Monte-Carlo move that integrates over
the total electric field $\bar{\bf{E}}$ (see also
Ref.~\onlinecite{acm3}).

\bibliography{mc}

\begin{thebibliography}{19}
\expandafter\ifx\csname natexlab\endcsname\relax\def\natexlab#1{#1}\fi
\expandafter\ifx\csname bibnamefont\endcsname\relax
  \def\bibnamefont#1{#1}\fi
\expandafter\ifx\csname bibfnamefont\endcsname\relax
  \def\bibfnamefont#1{#1}\fi
\expandafter\ifx\csname citenamefont\endcsname\relax
  \def\citenamefont#1{#1}\fi
\expandafter\ifx\csname url\endcsname\relax
  \def\url#1{\texttt{#1}}\fi
\expandafter\ifx\csname urlprefix\endcsname\relax\def\urlprefix{URL }\fi
\providecommand{\bibinfo}[2]{#2}
\providecommand{\eprint}[2][]{\url{#2}}

\bibitem[{\citenamefont{Schlick et~al.}(1998)\citenamefont{Schlick, Skeel,
  Brunger, Kal{\'e}, {Board Jr.}, Hermans, and Schulten}}]{schlick}
\bibinfo{author}{\bibfnamefont{T.}~\bibnamefont{Schlick}},
  \bibinfo{author}{\bibfnamefont{R.~D.} \bibnamefont{Skeel}},
  \bibinfo{author}{\bibfnamefont{A.~T.} \bibnamefont{Brunger}},
  \bibinfo{author}{\bibfnamefont{L.~V.} \bibnamefont{Kal{\'e}}},
  \bibinfo{author}{\bibfnamefont{J.~A.} \bibnamefont{{Board Jr.}}},
  \bibinfo{author}{\bibfnamefont{J.}~\bibnamefont{Hermans}}, \bibnamefont{and}
  \bibinfo{author}{\bibfnamefont{K.}~\bibnamefont{Schulten}},
  \bibinfo{journal}{J. Comp. Phys.} \textbf{\bibinfo{volume}{151}},
  \bibinfo{pages}{9} (\bibinfo{year}{1998}).

\bibitem[{\citenamefont{Sagui and Darden}(1999)}]{sagui}
\bibinfo{author}{\bibfnamefont{C.}~\bibnamefont{Sagui}} \bibnamefont{and}
  \bibinfo{author}{\bibfnamefont{T.}~\bibnamefont{Darden}},
  \bibinfo{journal}{Annu. Rev. Biophys. Biomol. Struct.}
  \textbf{\bibinfo{volume}{28}}, \bibinfo{pages}{155} (\bibinfo{year}{1999}).

\bibitem[{\citenamefont{Perram et~al.}(1988)\citenamefont{Perram, Petersen, and
  de~Leeuw}}]{ewald32}
\bibinfo{author}{\bibfnamefont{J.~W.} \bibnamefont{Perram}},
  \bibinfo{author}{\bibfnamefont{H.~G.} \bibnamefont{Petersen}},
  \bibnamefont{and} \bibinfo{author}{\bibfnamefont{S.~W.}
  \bibnamefont{de~Leeuw}}, \bibinfo{journal}{Molecular Phys.}
  \textbf{\bibinfo{volume}{65}}, \bibinfo{pages}{875} (\bibinfo{year}{1988}).

\bibitem[{\citenamefont{Beckers et~al.}(1988)\citenamefont{Beckers, Lowe, and
  de~Leeuw}}]{ewald}
\bibinfo{author}{\bibfnamefont{J.~V.~L.} \bibnamefont{Beckers}},
  \bibinfo{author}{\bibfnamefont{C.~P.} \bibnamefont{Lowe}}, \bibnamefont{and}
  \bibinfo{author}{\bibfnamefont{S.~W.} \bibnamefont{de~Leeuw}},
  \bibinfo{journal}{Molecular Simulation} \textbf{\bibinfo{volume}{20}},
  \bibinfo{pages}{269} (\bibinfo{year}{1988}).

\bibitem[{\citenamefont{Hockney and Eastwood}(1988)}]{ppm}
\bibinfo{author}{\bibfnamefont{R.~W.} \bibnamefont{Hockney}} \bibnamefont{and}
  \bibinfo{author}{\bibfnamefont{J.~W.} \bibnamefont{Eastwood}},
  \emph{\bibinfo{title}{Computer Simulation using Particles}}
  (\bibinfo{publisher}{Adam Hilger}, \bibinfo{year}{1988}).

\bibitem[{\citenamefont{Barnes and Hut}(1986)}]{multipole}
\bibinfo{author}{\bibfnamefont{J.~E.} \bibnamefont{Barnes}} \bibnamefont{and}
  \bibinfo{author}{\bibfnamefont{P.}~\bibnamefont{Hut}},
  \bibinfo{journal}{Nature} \textbf{\bibinfo{volume}{324}},
  \bibinfo{pages}{446} (\bibinfo{year}{1986}).

\bibitem[{\citenamefont{Essmann et~al.}(1995)\citenamefont{Essmann, Perera,
  Berkowitz, Darden, Lee, and Pedersen}}]{darden}
\bibinfo{author}{\bibfnamefont{E.}~\bibnamefont{Essmann}},
  \bibinfo{author}{\bibfnamefont{L.}~\bibnamefont{Perera}},
  \bibinfo{author}{\bibfnamefont{M.~L.} \bibnamefont{Berkowitz}},
  \bibinfo{author}{\bibfnamefont{T.}~\bibnamefont{Darden}},
  \bibinfo{author}{\bibfnamefont{H.}~\bibnamefont{Lee}}, \bibnamefont{and}
  \bibinfo{author}{\bibfnamefont{L.~G.} \bibnamefont{Pedersen}},
  \bibinfo{journal}{J. Chem. Phys.} \textbf{\bibinfo{volume}{103}},
  \bibinfo{pages}{8577} (\bibinfo{year}{1995}).

\bibitem[{\citenamefont{Maggs and Rossetto}(2002)}]{acm}
\bibinfo{author}{\bibfnamefont{A.~C.} \bibnamefont{Maggs}} \bibnamefont{and}
  \bibinfo{author}{\bibfnamefont{V.}~\bibnamefont{Rossetto}},
  \bibinfo{journal}{Phys. Rev. Lett.} \textbf{\bibinfo{volume}{88}},
  \bibinfo{pages}{196402} (\bibinfo{year}{2002}).

\bibitem[{\citenamefont{Maggs}(2002)}]{acm2}
\bibinfo{author}{\bibfnamefont{A.~C.} \bibnamefont{Maggs}},
  \bibinfo{journal}{J. Chem. Phys.} \textbf{\bibinfo{volume}{117}},
  \bibinfo{pages}{1975} (\bibinfo{year}{2002}).

\bibitem[{\citenamefont{Maggs}(2003)}]{acm3}
\bibinfo{author}{\bibfnamefont{A.~C.} \bibnamefont{Maggs}},
  \bibinfo{journal}{e-print cond-mat/0304521}  (\bibinfo{year}{2003}).

\bibitem[{\citenamefont{Schwinger et~al.}(1998)\citenamefont{Schwinger, Deraad,
  Milton, and yang Tsai}}]{schwinger}
\bibinfo{author}{\bibfnamefont{J.~S.} \bibnamefont{Schwinger}},
  \bibinfo{author}{\bibfnamefont{L.~L.} \bibnamefont{Deraad}},
  \bibinfo{author}{\bibfnamefont{K.~A.} \bibnamefont{Milton}},
  \bibnamefont{and} \bibinfo{author}{\bibfnamefont{W.}~\bibnamefont{yang
  Tsai}}, \emph{\bibinfo{title}{Classical Electrodynamics}}
  (\bibinfo{publisher}{Perseus Books}, \bibinfo{year}{1998}).

\bibitem[{\citenamefont{Sagui and Darden}(2001)}]{darden2}
\bibinfo{author}{\bibfnamefont{C.}~\bibnamefont{Sagui}} \bibnamefont{and}
  \bibinfo{author}{\bibfnamefont{T.}~\bibnamefont{Darden}},
  \bibinfo{journal}{J. Chem. Phys.} \textbf{\bibinfo{volume}{114}},
  \bibinfo{pages}{6578} (\bibinfo{year}{2001}).

\bibitem[{\citenamefont{de~Leeuw
  et~al.}(1980{\natexlab{a}})\citenamefont{de~Leeuw, Perram, and
  Smith}}]{leeuw1}
\bibinfo{author}{\bibfnamefont{S.~W.} \bibnamefont{de~Leeuw}},
  \bibinfo{author}{\bibfnamefont{J.~W.} \bibnamefont{Perram}},
  \bibnamefont{and} \bibinfo{author}{\bibfnamefont{E.~R.} \bibnamefont{Smith}},
  \bibinfo{journal}{Proc. Royal Soc. London} \textbf{\bibinfo{volume}{A373}},
  \bibinfo{pages}{26} (\bibinfo{year}{1980}{\natexlab{a}}).

\bibitem[{\citenamefont{de~Leeuw
  et~al.}(1980{\natexlab{b}})\citenamefont{de~Leeuw, Perram, and
  Smith}}]{leeuw2}
\bibinfo{author}{\bibfnamefont{S.~W.} \bibnamefont{de~Leeuw}},
  \bibinfo{author}{\bibfnamefont{J.~W.} \bibnamefont{Perram}},
  \bibnamefont{and} \bibinfo{author}{\bibfnamefont{E.~R.} \bibnamefont{Smith}},
  \bibinfo{journal}{Proc. Royal Soc. London} \textbf{\bibinfo{volume}{A373}},
  \bibinfo{pages}{57} (\bibinfo{year}{1980}{\natexlab{b}}).

\bibitem[{\citenamefont{Fraser et~al.}(1996)\citenamefont{Fraser, Foulkes,
  Rajagopal, Needs, Kenny, and Williamson}}]{fraser}
\bibinfo{author}{\bibfnamefont{L.~M.} \bibnamefont{Fraser}},
  \bibinfo{author}{\bibfnamefont{W.~M.~C.} \bibnamefont{Foulkes}},
  \bibinfo{author}{\bibfnamefont{G.}~\bibnamefont{Rajagopal}},
  \bibinfo{author}{\bibfnamefont{R.~J.} \bibnamefont{Needs}},
  \bibinfo{author}{\bibfnamefont{S.}~\bibnamefont{Kenny}}, \bibnamefont{and}
  \bibinfo{author}{\bibfnamefont{A.~J.} \bibnamefont{Williamson}},
  \bibinfo{journal}{Phys. Rev. B} \textbf{\bibinfo{volume}{53}},
  \bibinfo{pages}{1814} (\bibinfo{year}{1996}).

\bibitem[{\citenamefont{Chaikin and Lubensky}(1995)}]{chaikin}
\bibinfo{author}{\bibfnamefont{P.~M.} \bibnamefont{Chaikin}} \bibnamefont{and}
  \bibinfo{author}{\bibfnamefont{T.~C.} \bibnamefont{Lubensky}},
  \emph{\bibinfo{title}{Principles of condensed matter physics}}
  (\bibinfo{publisher}{Cambridge University Press}, \bibinfo{year}{1995}).

\bibitem[{\citenamefont{Deserno and Holm}(1998)}]{holm2}
\bibinfo{author}{\bibfnamefont{M.}~\bibnamefont{Deserno}} \bibnamefont{and}
  \bibinfo{author}{\bibfnamefont{C.}~\bibnamefont{Holm}}, \bibinfo{journal}{J.
  Chem. Phys.} \textbf{\bibinfo{volume}{109}}, \bibinfo{pages}{7678}
  (\bibinfo{year}{1998}).

\bibitem[{\citenamefont{Pais}(2002)}]{pais}
\bibinfo{author}{\bibfnamefont{A.}~\bibnamefont{Pais}},
  \emph{\bibinfo{title}{Inward bound. Of matter and forces in the physical
  world.}} (\bibinfo{publisher}{Oxford}, \bibinfo{year}{2002}).

\bibitem[{\citenamefont{Caillol}(1994)}]{caillol}
\bibinfo{author}{\bibfnamefont{J.-M.} \bibnamefont{Caillol}},
  \bibinfo{journal}{J.~Chem.~Phys.} \textbf{\bibinfo{volume}{101}},
  \bibinfo{pages}{6080} (\bibinfo{year}{1994}).

\end{thebibliography}

\end{document}